\documentclass[aps,showpacs]{revtex4}

\usepackage{fullpage,amsfonts,amssymb,pst-node,epsfig}

\begin{document}
\title{Coherent bubble-sum approximation for coupled-channel
resonance scattering}
\author{N.E. Ligterink}
\affiliation{
Department of Physics and Astronomy, University of
Pittsburgh, \\
3941 O'Hara Street, Pittsburgh, PA 15260, U.S.A.}
\begin{abstract}
For coupled-channel resonance scattering we derive a model
with a
closed form solution for the $T$-matrix that satisfies 
unitarity and analyticity. The two-channel case is
handled explicitly for an arbitrary number of resonances.
The method focuses on the expansion of the transition matrix
elements, $\Gamma(s)$, in known analytical functions.
The appropriate hadronic form factors and the related 
energy shifts can be determined from the scattering data.
The differences between this method and the $K$-matrix
and the Breit-Wigner approximation are illustrated in 
the case of the $S_{11}$ resonances $S_{11}(1535)$ and 
$S_{11}(1650)$.
\end{abstract}
\pacs{24.30.-v,
24.10.Eq,
11.80.Gw,
11.55.-m
}

\maketitle

\section{Introduction}

Analyticity is numerically the hardest problem to keep
track of in a scattering problem. Both the singularities
from on-shell states, which yield imaginary parts and
the effects of thresholds require careful analysis. This
is one of the main hurdles for the development of unitary
models for coupled-channel scattering. 

In this short paper a set of consistent approximations that yield
unitary and analytical results for coupled channel problems
is described. It is accompanied by the open source program 
BUBBLEGUM, which yield numerical $T$-matrix results, together
with the corresponding $K$-matrix approximation, \cite{New66}
Breit-Wigner approximation, \cite{Fes92,Fel69}
and the perturbative results. The general purpose of both 
the paper and the computer program is threefold. First, to show that
approximations that violate unitarity or analyticity will 
generally deviate considerably \cite{BH98}
from exact results in the case of strong interactions
and thresholds close to resonance values. Second, to show
that the exact results can be derived straightforwardly
with a reasonable effort, in a manner that can improved be
systematically. Third, to show that scattering data do not
only contain pole information of the resonances but also
the non-perturbative matrix elements between asymptotic,
or scattering states, and resonances. Extraction of these
matrix elements from the scattering data
might lead to better constraints on models for hadrons.

Furthermore, 
substantial energy shifts occur through coupling 
of resonances with asymptotic states, or decay channels,
which can only be analyzed in a fully analytic approach.
Since hadrons are composite particles, they have form
factors that suppress high-energy processes. These
form factors and the energy shifts are related and
should be treated consistently, which neither the $K$-matrix
approach nor the Breit-Wigner approximation can do.
Depending on the large energy behavior of the form factor, the shift
can vary substantially. In the case of suppression at
high energy, the shift will change sign as scattering energy
increases, from negative to positive.

It is common folkore that bare masses have no meaning. 
It stems from renormalizable field theory, where indeed masses
can undergo abritrary shifts through finite renormalization.
However, in hadron theory constituent quark models and
other hadronic models yield the bare states with
the bare masses and their couplings to the decay channels.
In this case a comparison between scattering data and
these models should incorporate the energy shifts through
the non-perturbative coupling of bare states and decay 
channels.

\section{Theory}

The simplest process contributing to $s$-channel scattering
is the tree diagram, where the two-particle state $j$, in a
particular partial wave, forms a resonance state $b$ which decays
into a two-particle state $i$. The transition amplitude is given
by:
\begin{equation}
T^{(0)}_{ij} = g_{ib} f_i(E) \sqrt{\rho_i(E)}
\frac{1}{E - M_b + i\epsilon}
\sqrt{\rho_j(E)} f_j(E)  g^\ast_{jb} \ \ ,
\label{eqtree}
\end{equation}
where $g_{ib}$ and $g_{jb}$ are the coupling constants,
$\rho(E)$ is the phase space for each state, $f(E)$ the form factors,
$M_b$ the mass of the state $b$, and $E$ the scattering energy.

This lowest order diagram is a poor approximation \cite{Dir30} to
the actual scattering process, especially for hadron dynamics. First,
the coupling constant is large, therefore, during scattering,
the initial state $j$ goes, on average, through several intermediate
states, before ending up in the final state $i$. Second, for scattering
energies $E$ close to the resonance mass $M$, the interaction is
strong due to the small energy denominators, even if the coupling is
small. Hence, the simple tree
diagram, which corresponds to the lowest order perturbative expansion,
fails here. If the problem would be solved consistently, the resonance
will acquire a width, which corresponds to the shift of the pole
into the complex plane.

Solving the scattering problem consistently means iterating the
elementary scattering processes such that it reaches a fixed point.
There are many more or less equivalent ways to find the self-consistent
solution. In this paper we will use the Lipmann-Schwinger equation
\cite{Fes92,GW64}:
\begin{equation}
T_{ca} = T^{(0)}_{ca}  + \sum_d T^{(0)}_{cd}
\frac{1}{E-E_d+i \epsilon} T_{da} \ \ ,
\label{eqls}
\end{equation}
which can be written down in many other forms. In practice, this
equation determines the transition amplitude $T$, which is the
sum over all possible processes, with different numbers of intermediate
states.

For the case of two coupled channels 
the matrix  Lipmann-Schwinger equation will look like:
\begin{equation}
\left( \begin{array}{cc} t_{11} & t_{12} \\ t_{21} & t_{22} \end{array}
\right) = X + X  \cdot
\left( \begin{array}{cc} i W_1 + \tilde W_1 & 0 \\
0 &  i W_2 + \tilde W_2 \end{array} \right) \cdot
\left( \begin{array}{cc} t_{11} & t_{12} \cr t_{21} & t_{22} \end{array}
\right) \ \ , \label{eqt22}
\end{equation}
where
\begin{equation}
X = \left( \sum_a \frac{1}{E-M_a}\left( \begin{array}{cc} |g_{1a}|^2 &
g_{1a}g_{2a}^\ast
\cr g_{1a}^\ast g_{2a} & |g_{2a}|^2
 \end{array} \right) \right) \ \ .
\end{equation}
The reduced transition amplitudes $t$ are defined as:
\begin{equation}
T_{ij}(E) = |i\rangle t_{ij}(E) \langle j| \ \ ,
\end{equation}
with the states $|i\rangle$ given by:
\begin{equation}
|i\rangle = \int d E \ \sqrt{\rho_i(E)} f_i(E) |\phi_i(E)\rangle \ \ .
\end{equation}
$ |\phi_i(E)\rangle$ is the asymptotic state $i$ with energy $E$.

The solution for this case of two coupled channels can still be
given in a simple closed form:
\begin{eqnarray}
t_{11} & = & \frac{X_{11} - {\rm det[X]} {\cal W}_2}{ 1 - X_{11} {\cal
W}_1 - X_{22} {\cal W}_2
+ {\rm det[X]} {\cal W}_1 {\cal W}_2} \ \ ,  \label{t11}\\
t_{22} & = & \frac{X_{22} - {\rm det[X]} {\cal W}_1}{
1 - X_{11} {\cal W}_1 - X_{22} {\cal W}_2
+ {\rm det[X]} {\cal W}_1 {\cal W}_2} \ \ ,  \label{t22}\\
t_{12} & = &  \frac{X_{12}}{1 - X_{11} {\cal W}_1 - X_{22} {\cal W}_2
+ {\rm det[X]}{\cal W}_1  {\cal W}_2} \ \ , \label{t12} \\
t_{21} & = & \frac{X_{21}}{1 - X_{11} {\cal W}_1 - X_{22} {\cal W}_2
+ {\rm det[X]} {\cal W}_1 {\cal W}_2} \ \ ,
\label{t21}
\end{eqnarray}
where ${\cal W}_j$ is defined as: ${\cal W}_j = i W_j(E) + \tilde
W_j(E)$, and the 
functions $W$ are defined through the following integrals:
\begin{eqnarray}
W_i(E) & = & \pi \int d E' \langle \phi_i(E')|\phi_i(E)\rangle
\rho_i(E) f_i(E)^2 =
\pi \rho_i(E) f_i(E)^2  \ \ , \label{eqW} \\
\tilde W_i(E) & = & {\rm PV} \int d E' \frac{\rho_i(E')
f_i(E')^2}{ E-E'} \ \ , \label{eqWt}
\end{eqnarray}
where PV stands for a principal value integral, and the asymptotic
states are orthonormal: $\langle \phi_i(E')|\phi_i(E)\rangle =
\delta(E'-E)$. It is clear that these results can easily be  extended
to incorporate an arbitrary number of channels.

Both $\rho$ and $f$ are real; any phase information
is to be incorporated in the coupling constants. In the absence of a
form factor $f$ the second integral generally does not converge. It
requires regularization, and possibly a motivation for this
regularization through renormalization. Subtracted dispersion integrals
effectively correspond to regularization through local subtractions. Another 
option is to
set $\tilde W$ to zero, and only use the phase space for the imaginary
part, which is known as the $K$-matrix method.

For a general case the dispersion integral Eq.~(\ref{eqWt}) cannot be solved
analytically. However, since the form factor is rather arbitrary, it can
be chosen such that $\rho_i(E) f_i(E)^2$ has a known solution
for the dispersion integral. An expansion in a set of these functions, 
allows one to determine $\rho(E) f(E)^2$ as
part of the analysis of scattering data of resonances. This quantity
can, and should, be compared to models of hadrons since the function 
$\rho(E)
|g f(E)|^2$ is the square of the transition matrix element of the
Hamiltonian:
\begin{equation}
\rho(E)|g f(E)|^2 = |\langle \phi(E) | H | M\rangle|^2\ \ ,
\end{equation}
where $M$ is the bare resonance state.
The transition matrix element is expected to fall-off at higher 
energies, due to the composite nature of hadrons. 

Another constraint on $W(E)$ is given by general covariance, which
requires that $W(E)$ is analytic in $s=E^2$ with all the cuts on
the real axis associated with decay channels, or asymptotic states.
In Feynman perturbation theory covariance is the result of
summing the different time-ordered diagrams. In the language of
states, these different time orderings correspond to different
states, which often correspond to decay channels that will
be of little significance in the kinematical region of interest.
However, as restoring covariance comes at little cost,
we replace the energy dispersion relation with:
\begin{equation}
\tilde W(E')=
\int d E \frac{W(E)}{E'-E} \ \ \to \ \
\int d s \frac{W(\sqrt{s})}{E'^2 - s} \ \ ,
\label{eqetos}
\end{equation}
would yield generally a small difference in $\tilde W$. This, however,
includes the second time ordering and restores analyticity in $s$,
which is a consequence of full covariance. 
Moreover, analyticity in $s$ implies a restriction on possible
approximations for $\tilde W(E)$, {\it i.e.}, there should not be a cut at $s=0$.
In $W(E)$ itself the second time ordering is not included, as it
would correspond to a different decay state which is generally of little
consequence, however, would yield additional singularities to
keep track of. This feature of manifest relativistic invariant formulation
makes the evaluation of higher order Feynman diagrams in Minkowski
space complicated. The advantage of the one-state-one-singularity approach
advocated here is that the singular structure of the perturbative kernel
of the Lipmann-Schwinger equation $T^{(0)}$ is near to trivial.
Characterizing states by their energy simplifies the equations to
such an extent that a greater part of the calculations can be
performed analytically, as we will see below.
Note, when the kinematical domain of validity is restricted 
to the energies where only the designated states, that appear
as channels in the Lipmann-Schwinger equation, can go on shell, the
result is fully covariant with the replacement Eq.~(\ref{eqetos}).

Given a particular threshold behavior $n$, associated with a partial wave,
we use as leading order approximation for $W(E)$:
\begin{equation}
W^{(0)}(\sqrt{s}) = \sqrt{s_{\rm th}}
\frac{(s/s_{\rm th} -1)^{n/2}}{(s/s_{\rm th})^{[n/2]+1}}
\theta(s/s_{\rm th} -1)\ \ ,
\end{equation}
where $s_{\rm th}$ is the threshold energy squared, and
$[i]$ is the integer part of $i$. The power $n$ is not only related
to the partial wave, but also to kinematical factors which are
different for a heavy-light system like $\pi N$ and an equal
mass system, like $\pi \pi$. Furthermore, it depends as well on
the kinematical range of the threshold, {\it i.e.}, if the threshold extends
to where the masses are comparable to the energy, it would yield
a different threshold behavior. For example, in the $\pi N$ scattering
in the $\Delta$ region, the real threshold behavior stops at about
50 MeV above the threshold.

An additional expansion to model or fit the function $W$ away from
threshold is given by a polynomial $P_m$ of order $m$ in 
$\xi = (s - s_{\rm th})/s$:
\begin{eqnarray}
W(\sqrt{s}) & = & \theta(s/s_{\rm th} -1)
\frac{s_{\rm th}^{[n/2]-n/2 + 3/2}}{s^{[n/2]-n/2 + 1}} 
\xi^{n/2} P_m(\xi) \label{eqp} \\
& = & \theta(s/s_{\rm th} -1)
\sqrt{s_{\rm th}} \sum_{i=0}^m c_i \frac{(s/s_{\rm th}
-1)^{n/2 + i}}{(s/s_{\rm th})^{[n/2]+ i + 1}}  \\
& \equiv & \theta(s/s_{\rm th} -1)
\sqrt{s_{\rm th}} \sum_{i=0}^m c_i w_{(n/2+i)([n/2]+i+1)}\ \ ,
\label{eqex}
\end{eqnarray}
such that $\tilde W$ is analytical in $s=0$ and has the dimension
of energy. The first coefficient $c_0=1$ such that the
threshold behavior is determined solely by the power $n$ and
the coupling constant. 
The coupling constants $g_{ib}$ are dimensionless. Note,
that no additional dimensionful quantities are introduced beside
the threshold energy, which makes a good candidate for a normalized
description of resonance scattering, which is up to now littered
with different form factors with all kind of dimensionful quantities
with little meaning. If any dimensional quantities, such as
the QCD scale, pion decay constant, or the pion mass, are 
relevant the can be expressed
as a number times the threshold energy $\sqrt{s_{\rm
th}}$.

Each of these terms in the polynomial expansion has a closed form
solution for the dispersion integral, where we distinquish  integer
and half-integer values for $n$ in $w_{nk}$:
\begin{eqnarray}
w_{nk} & = & \theta(\hat s -1) \pi \frac{(\hat s-1)^n}{\hat s^k} \\
\tilde w_{nk} & = & 
(1-{\cal T}_0)\frac{(\hat s-1)^n\log|1-\hat s|}{\hat s^k}  \\
w_{\frac{2n+1}{2}k} & = & \pi\frac{{\rm Re}(\hat s-1)^{(2n+1)/2}}{\hat s^k} \\
\tilde w_{\frac{2n+1}{2}k} & = & (-1)^{n}(1-{\cal T}_0)
\pi\frac{{\rm Re}(1-\hat s)^{(2n+1)/2}}{\hat s^k}
\end{eqnarray}
where ${\cal T}_0$ refers to a Taylor-Laurent expansion of all the
singular terms in $s^{-1}$, and $\hat s = s/s_{\rm th}$.
The function $\xi^m/\hat s$ Eq.~(\ref{eqp}) peaks at 
$ s = (m+1) s_{\rm th} $, while $\xi^m/\sqrt{\hat s}$
peaks at $s = (2m+1) s_{\rm th} $, such that
every higher order term in $P_m(\xi)$ extend the range of energies
over which $W(E)$ can be fitted. The
polynomials form a $L^1$ basis for $W$ on $0<1/(1+\xi)<1$.

The real part $\tilde w$ can be derived in closed form
by expanding the series and resumming the terms, analogous to
the calculation in \cite{Lig00}. However, since the singularity
at $s=0$ has to cancel, and the difference can only be meromorphic
functions that fall off at infinity, it is simple to see that
the real part is restricted
to the trivial analytical continuation and the Laurent series.
The functions $\tilde w$ generally have a large negative value,
leading to a negative mass shift. This is mainly due to the large
high-energy tail $\tilde w_{\frac{2n+1}{2}k} \sim 1/\sqrt{s} $ or 
$\tilde w_{nk} \sim 1/s$. This leading order behavior is necessary
to be able to fit arbitrary $W(E)$ dependence, however, might be 
suppressed in the actual scattering, with an equal reduction
of the mass shift.

Using these functions to solve the Lipmann-Schwinger equation in closed
form effectively means making a simultaneous, or coherent, bubble sum
approximation in each of the decay channels for each of the resonances.
Therefore we refer to this method as the Coherent Bubble Sum Approximation.
Single bubble sum approximations have been used widely. It
corresponds to dressing the propagator of the resonance with 
its single or multiple decay channels. However, the case of an unitary, 
analytical, and covariant approach with channel mixing has only
appeared at the heart of a few coupled-channel analyses with fixed
form factors. \cite{CFHK79,VDL,Surya} 
In the case of a single resonance, the bubble sum can be seen as a dressing
of the resonance particle; a change in the particle properties. In
the case of coupled channels such an interpretation does not
exist. In this case only the physical observables in the scattering
experiment are free of ambiguities in the interpretation. In simple 
terms, the resonance properties are intertwined and co-dependent.

Solving the Lipmann-Schwinger equation Eq.~(\ref{eqls}) reduces
to solving the algebraic equation Eq.~(\ref{eqt22}), and determining
the Laurent-Taylor expansions, which is automated in the BUBBLEGUM
code.

\section{Mass shifts}

The $K$-matrix method leaves the real part of the resonance pole in
the same position. The coherent bubble sum shifts the pole in 
order to satisfy analyticity contraints. These shifts can be
substantial. However, the word ``mass shifts'' suggest that all
effects can be reabsorbed in shifting the mass such that it
accounts for main differences between $K$-matrix and $T$-matrix results.
This is generally not the case. In a coupled channel problem, the
mass shifts differ from channel to channel. In the $K$-matrix
the real part of the pole, {\it i.e.}, the position where the real
part of the amplitude crosses the axis, is at the same location in every
channel as long as the coupling constant is not too large. 
Only for large coupling constants unitary yields stringent constraints
for the $K$-matrix, which can shift theses locations independently. 
For the exact result this does not hold. The isobar
model, \cite{Fel69} which implies resonance properties independent of
the decay channels, is therefore flawed from the start.
The procedure of finite mass renormalization is not unique.

However, in order to make the comparison between $K$-matrix
and exact results more meaningful for the $K$-matrix method, one can
make a global mass shift $\Delta M_a$ for resonance $a$, 
which, in some way, averages the shifts in each of the channel:
\begin{equation}
\Delta M_a = \lim_{E\to M_a}\frac{(E-M_a)
(1- X_{11} {\cal W}_1 - X_{22} {\cal W}_2 + 
{\rm det[X]} {\cal W}_1 {\cal W}_2)}
{\frac{\partial}{\partial M_a} (E-M_a)
(1- X_{11} {\cal W}_1 - X_{22} {\cal W}_2 +
{\rm det[X]} {\cal W}_1 {\cal W}_2)} \ \ ,
\label{eqms}
\end{equation}
where the expression is given by the common denominator
in all of the $T$-matrix channels, Eq.~(\ref{t11}-\ref{t21}).
The term in the denominator corresponds to the appropriate
wave function renormalization. 
In the case of small coupled-channel effects and small
overlap of the resonances the real part of the shifted
pole positions coincide with the expected channel-independent
pole position. However, in the case large couplings or
coupled-channel effects this independent resonance 
mass shift breaks down, {\it i.e.}, linear approximation of 
the denominator around the pole is no longer valid.
In that case an iterative optimizing scheme to recover the mass
shifts is required. However, as the mass shift is not
an invariant quantity, the results from such a scheme 
would have no real significance. Eventually one has to 
accept that the scattering data and the bare properties
and couplings are the only well-defined quantities. 
However, the bare masses must always be given in 
combination with the coupling constants and the form 
factors.

\section{$t$-exhange without open channels}

This method of comparing scattering data to resonance properties can
also serve as an approximate method for handling additional $t$-exchange
processes. For low and medium energy scattering, $t$-exchange diagrams
will not go on-shell, {\it i.e.}, there is no imaginary part associated with 
the perturbative diagram $T_{\rm exch}^{(0)}$.  However, the real
part is a direct consequence of the energies at which the $t$-exchange
diagram goes on-shell and strength in the channel.
Therefore, the contributions can be approximated by a single resonance:
\begin{eqnarray} 
{\rm Im} T_{\rm exch}^{(0)} & =  & -\int \mu(E') d E'
g_{ix} f_i(E,E') \sqrt{\rho_i(E)}
\pi \delta(E - E')
\sqrt{\rho_j(E)} f_j(E,E')  g^\ast_{jx}  \nonumber  \\
& \to &
 - \overline{\mu} g_{ix} f_i(E,M_{\rm eff}) \sqrt{\rho_i(E)}
\pi \delta(E - M_{\rm eff})
\sqrt{\rho_j(E)} f_j(E,M_{\rm eff})  g^\ast_{jx}\ \ , \\
T_{\rm exch}^{(0)} & =  & \overline{\mu} g_{ix} f_i(E,M_{\rm eff})
\sqrt{\rho_i(E)} \frac{1}{ E - M_{\rm eff} + i \epsilon}
\sqrt{\rho_j(E)} f_j(E,M_{\rm eff})  g^\ast_{jx}\ \ ,
\label{eqtexch}
\end{eqnarray}
where $\mu(E')$ is a measure of the kinematical range over which the
$t$-exchange diagram can go on shell, and $M_{\rm eff}$ is the mean
of that measure:
\begin{eqnarray}
\int \mu(E') d E' & = & \overline{\mu} \ \ , \\
\int \mu(E') d E' E' & = & M_{\rm eff} \ \ .
\end{eqnarray}
In the case where the $t$-exchange diagram has an imaginary part
in the kinematical range of interest, this approximation will not
hold. More appropriate methods  for this problem are under 
investigation and will be reported later.

In effective field theory highly virtual intermediate states
are replaced by effective contact interactions. The effective
resonance described here does more than that. The effective
resonance that replaces the virtual $t$-exchange incorporates
also the leading energy dependence of the $t$-exchange contribution. 

\section{$S_{11}$ resonances}

\begin{figure}
\centerline{\includegraphics[width=14cm]{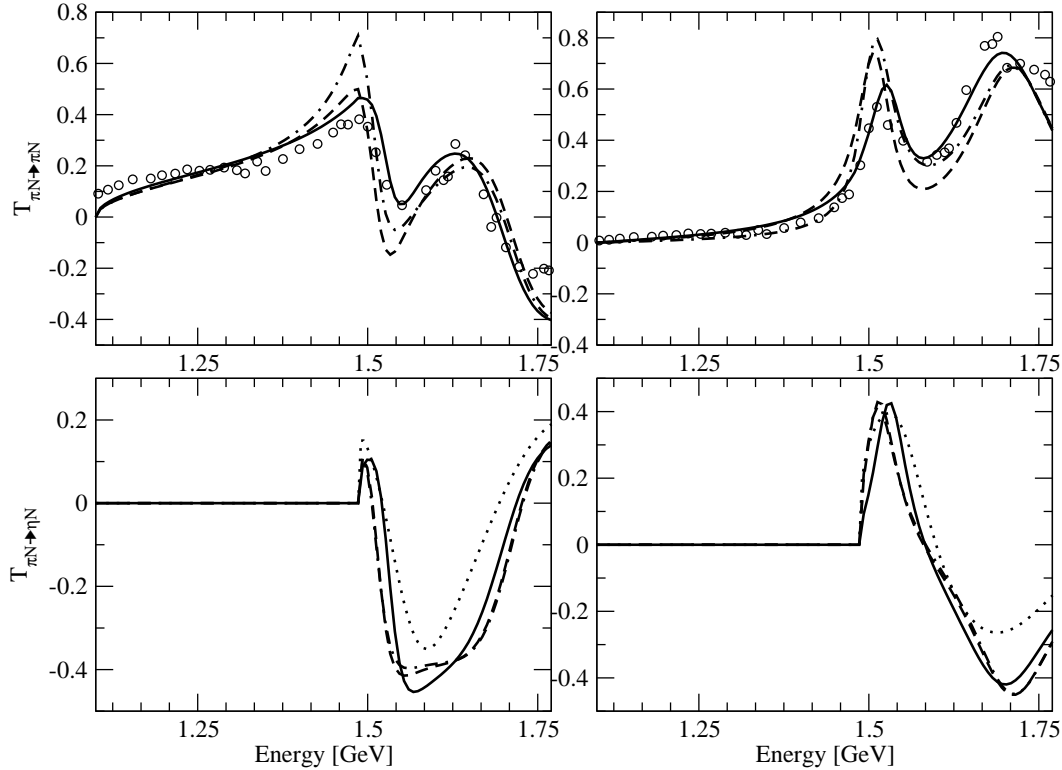}}
\caption{The transition amplitudes in $S_{11}$ $\pi N$
scattering. The top row are the real and imaginary part
of the $\pi N \to \pi N$ amplitudes, and the bottom row
the $\pi N \to \eta N$ amplitudes.
The dots and circles are partial wave data, without error bars,
the solid lines are the $T$-matrix results, the dashed lines
the $K$-matrix approximation of these results, with identical
parameters, and the dot-dashed lines the Breit-Wigner results,
as described in the paper.}\label{fig}
\end{figure}

The $S_{11}$ resonances at 1.535 GeV and 1.65 GeV are a good test case
for the coupled channel analysis of overlapping resonances as
the threshold of the $\eta N$ channel is in the same region.
However, there is only limited data available for the $\eta N$ channel.
We fit the data between threshold and 1.77 GeV with minimal model of 
two bare masses and four coupling constants. Beyond the resonance
region there is still a substantial amplitude, and little data in
other than the $\pi N$ channel, which might indicate the
importance of $\pi \pi N$ 
final states, and $t$-channel exchanges. A kink in the 
$\pi N \to \pi N$ data at 1.3 GeV is a further indication
of $\pi \pi N$ states. For the clarity of the
example we do not include the high-energy  region and work with 
the minimal parameter set of six parameters. We use the 
GWU-VPI partial wave data\cite{ASWP}
for the $\pi N$ elastic amplitude and the results from the
Pitt-ANL analysis\cite{VDL} for the $\pi N \to \eta N$ channel.
Standard fitting algorithms do not work properly as the 
variables are highly correlated in a non-trivial manner due to
the unitarity and analyticity conditions. Instead we performed a
global search, with an increasing mesh, zooming in on the optimal
fit. 

In Figure~\ref{fig} we compare the exact $T$-matrix results with
the $K$-matrix and Breit-Wigner approximations. The top left 
figure shows real part of the $\pi N \to \pi N$ transition amplitude,
while the imaginary part appears on the right. The bottom row
is the same combination for the $\pi N \to \eta N$ transition
amplitude. The bare masses are 1.60 GeV and 1.79 GeV, while the
renormalized masses come to 1.51 GeV and 1.71 GeV. The latter 
are used for the $K$-matrix and the Breit-Wigner approximations.
Without these mass shifts Eq.~(\ref{eqms}), the approximate 
results would have no resemblance with the data.
The dotted line and the circles are the Pitt-ANL and GWU-VPI partial waves 
for the $\eta N$ and the $\pi N$ channels respectively. The deviation
for the $\eta N$ partial wave is due to the large error uncertainty,
which led to the dominance of the $\pi N$ channel in the fit. However,
varying the relative importance of the $\eta N$ data, with
respect to the $\pi N$ data, did not significantly
altered the results. It seems that the Pitt-ANL and GWU-VPI partial 
wave analysis, which include background terms, lead to an
inconsistent $\eta N$ amplitude that cannot be fitted with
with a resonance-only model like the coherent bubble sum 
approximation. With the current $\eta N$ data it is not possible
to determine the nature of the deviations, and whether the data
favors a background contribution. However, there seems to be
no natural explanation for a background contribution. 
The background should model some degrees of freedom, or states,
that are important yet virtual, {\it i.e.}, not associated with open 
channels, such as, perhaps, the $\rho$-exchange between the $\pi N$
pair.

Including the first term $c_1$ in the expansion Eq.~(\ref{eqex})
of the form factors halves the mean square deviation between
the data and the fit. Notably,
it increases the effective width of the $\pi N$ form factor,
while it decreases the effective width of the $\eta N$
form factors, corresponding, respectively, to a positive
and a negative expansion coefficient $c_1$.

Note that it would be possible to obtain better fits in the 
$K$-matrix approximation and the Breit-Wigner approximation.
However, the purpose of this paper is to point out the 
differences which are
solely the result of the approximations made, keeping the
model identical. In this case it is clear that the first
resonance $S_{11}(1535)$ is overestimated in both
approximations. The coupling constants are $g_{1,\pi N}=-0.11$,
$g_{2,\pi N} = 0.175$, $g_{1,\eta N}= -0.10$, and
$g_{2,\eta N}= -0.11$.

\section{conclusions}

The $K$-matrix approximation does not require the solution
of the dispersion integral. Therefore, it is often used to
incorporate more complex interactions. In the Coherent Bubble 
Sum Approximation it correspond to setting $\tilde W_i$ to
zero. The Breit-Wigner approximation is generally not a
well-defined procedure. Eventually it boils down to replacing
the complex interaction by a sum of Breit-Wigner forms with partial
widths for each of the decay channels in each of the resonances:
\begin{equation}
T_{ij} = \sum_b \frac{ g_{ib}\sqrt{\Gamma_i(E) \Gamma_j(E)}g_{jb}^\ast}
{E- M_b - \sum |g_{ib}|^2\Gamma_i(E)} \ \ ,
\end{equation}
where, in principle, each of the coupling constants and $\rho_i$'s 
are unknown and 
should be recovered from the analysis. However, given the 
$W(E)$'s and the coupling constants, the closest general analogy is
given by $\Gamma(E) = W_i(E)$. From the example of the $S_{11}$ 
resonances, where the coupling strength is moderate, it is
clear that both the $K$-matrix and the Breit-Wigner approximation
leads to significant deviations in the transition amplitudes.

A feature of scattering data where approximate methods such as
Breit-Wigner and $K$-matrix approximations yield erroneous results
is the interaction between thresholds and subthreshold resonances.
A resonance with a strong coupling to a channel with a threshold
at a higher energy might lead to anomalous threshold behavior.
Once a resonance has an energy close enough to the threshold energy,
it might get attracted into the channel, giving either a full
circle in the Argand plot, or a sharp spike in the real part
of the transition amplitude. Neither feature is properly 
reproduced in the Breit-Wigner or the $K$-matrix method. 

In this paper the form of $W(E)$ is not derived. Only simple 
assumptions about its threshold behavior are made. The
energy dependence can be modeled by a polynomial in the
variable $(s-s_{\rm th})/s$. This freedom is an essential
part of the Coherent Bubble Sum Approximation. If possible,
the data should determine the function $W(E)$. Its value
should be extracted from the scattering data without model
dependence. It was only assumed that $W(E)$ falls off at
infinity, which is considered realistic for composite hadronic
systems. Analyzing resonance scattering data and modeling
transition matrix elements of hadronic states are two
separate problems and should be treated as such. In 
future studies we will examine $t$-exchange, multi-particle
final states, and field-theoretical approaches which 
require renormalization.

\end{document}